\documentclass[12pt]{article}
\usepackage[dvips]{graphicx}
\usepackage{amssymb}

\renewcommand{\(}{\begin{equation}}
\renewcommand{\)}{\end{equation}}

\newcommand{\ds}{\displaystyle}
\newcommand{\barray}[1]{\begin{array}{#1}}
\newcommand{\earray}{\end{array}}
\newcommand{\ket}{\rangle}
\newcommand{\bra}{\langle}
\newcommand{\de}{\partial}
\newcommand{\desq}{\partial^2\!}

\newcommand{\ov}[1]{\overline{#1}}

\makeatletter
\newbox\unobox 
\setbox\unobox=\hbox{\kern+.3em {\rm R}%
\kern-0.88em {\rm I\kern+0.9em}}
\def\erreuno{\copy\unobox} 
\newbox\duebox 
\setbox\duebox=\hbox{\kern+.3em {\rm R}%
\kern-0.88em {\rm I}%
\kern+.5em\raise.88ex\hbox{\scriptsize 2}\kern+.3em}
\def\erredue{\copy\duebox} 
\def\tfract#1/#2{{\textstyle{\raise0.8pt\hbox{$\scriptstyle#1$}\over%
\hbox{\lower0.8pt\hbox{$\scriptstyle#2$}}}}}

\begin{document}

\begin{titlepage} 
\begin{flushright}
 IFUP-TH. 26/2000\\
  July  2000\\
\end{flushright}

\vspace{8mm}

\begin{center}

{\Large \bf Statistical properties of classical  

\vskip 0.5truecm

gravitating particles in (2+1) dimensions}

\vskip 1.7 truecm 

\vspace{12mm}
{\large \bf M. ~Ghilardi ${}^{a}$ ~and~  E. ~Guadagnini
${}^{b}$}

\vskip 0.8 truecm 

\vspace{6mm}
($a$) Scuola Normale Superiore di Pisa \\ 
Piazza dei Cavalieri, 7. $\; $ I-56100 PISA. Italy

\vspace{3mm}
($b$) Dipartimento di Fisica dell'Universit\`a di Pisa \\ 
(and INFN , sezione di Pisa) \\
Via F. Buonarroti, 2. $\; $ I-56100 PISA. Italy 

\vspace{6mm}
E-Mail: ghilardi@cibs.sns.it ~,  $\;$ guada@df.unipi.it 

\vspace{3mm}

\end{center}
\vspace{10mm}

\begin{quote}
\hspace*{5mm} {\bf Abstract.}  ~We report the statistical properties of classical 
particles in (2+1) gravity as resulting from numerical simulations. 
Only particle momenta have been taken into account. In the range of total momentum
where thermal equilibrium is reached, the distribution function and
the corresponding Boltzmann entropy are computed. In the presence of large gravity effects, 
different extensions of the temperature turn out to be inequivalent, the distribution function has
a power law high-energy tail and the entropy as a function of the internal energy presents a flex. 
When the energy approaches the open universe limit, the entropy and the mean value of the particle
kinetic energy  seem to diverge.  

\end{quote}
\end{titlepage}
\clearpage

\noindent {\bf 1. Introduction.} ~The inclusion of gravitational effects in
thermodynamics has important consequences at the cosmological level.  There is
a particular interest in this subject also because gravity deeply modifies
certain basic features of non-gravitating systems; for instance,  in the presence
of gravitational interactions one cannot define the thermodynamic limit in the standard
way, one may find  bounds for the value of the entropy, the additivity of the
thermodynamic potentials is no longer valid in general, etc.  In this article we 
explore the statistical properties of a classical gravitating gas in (2+1) dimensions
with vanishing cosmological constant. The gravitational dynamics in (2+1) dimensions is
rather simple but it is not trivial, so this is one of the simplest models in which one
can look for new phenomena.  

Our results are based on numerical simulations. We have considered a set of
classical pointlike particles, with random initial momenta, moving  on a  spatial surface of
topological type either \erredue or the sphere $ \, {\rm S}^2 \, $ for a finite  time interval. 
We followed only the evolution of the particle momenta.  Let $ \, G \, $ be the gravitational
constant and  $ \, U \, $  the total energy of the system. When $ \, 4 G U <  1 \, $, the system
reaches thermal equilibrium. Whereas, for space-like total momentum or when $ \, 4 G U >  1 \, $,
the numerical analysis shows the presence of dynamical instabilities; jets of particles appear with
no-limits increasing  energy. 

In the case $ \, 4 G U <  1 \, $, the resulting distribution function in momentum space  $ \, f \,
$ is numerically produced and the corresponding Boltzmann entropy, $ \, S \propto \int f \, \log f
\, $, is computed as a function of the total energy $ \, U \, $ of the system.  If  $ \, 4 G U \ll
1 \, $ the gravitational effects  can be neglected and the entropy $ \, S \, $ shows the classical
$ \, \log U \, $ behaviour. On the other hand, when the value of the energy approaches 
the limit   $ \, 4 G U = 1 \, $, gravity has a strong influence on the dynamics of the
particles and the value of $ \, S \, $  seems to diverge.  The implications of this behaviour on
the heat capacity of the gas are considered. Finally, we compare  various definitions of
temperature which turn out to be inequivalent in the energy range of large gravity effects.  

\bigskip 

\noindent{\bf 2. Classical gravitating particles.}  ~The peculiar properties of (2+1)
gravity have been described in Ref. [1,-,4]. In any (2+1) spacetime $ \, M \, $ 
containing classical pointlike particles, all the curvature is located on the 
worldlines of the particles. Each particle moves in a locally flat surrounding
spacetime. Since the isometry group of flat spacetime is isomorphic with the Poincar\'e
group $ISO(2,1)$, both the kinematics and the laws of the dynamics  are covariant  with
respect to transformations of the Lorentz group $SO(2,1)$.  

A rather simple description of the spacetimes containing $ \, N \, $ pointlike
gravitating particles  has been produced by 't~Hooft [5,6]. This description is given by
the linear evolution of Cauchy surfaces  which are tiled
by spatial planar polygons; the time evolution also includes modifications of the tiling
combinatorics according to codified transition rules [6]. A Lorentz frame is associated
to each polygon with local coordinates $ \, (t, x, y) \, $ and metric  $\, ds^2 = dt^2 -
dx^2 - dy^2 \, $; with respect to this frame, the interior of each polygon belongs to the
spatial $ \, t=$ constant surface.   The extrinsic curvature is vanishing in the
interior of each tile and is singular on the edges. Each particle is placed in one
corner of a polygon; this corner is the intersection point of two consecutive edges which
must be identified.     

At a given initial time $ \, t = t_0 \, $, one can always introduce a tessellation in
which all the particles are placed at the corners of one polygon $ \, {\cal P}(t_0) \,
$; let us denote by  $ \, LS \, $  the  Lorentz system which is associated with  $ \,
{\cal P} (t_0) \, $. The initial values $ \, \{ \, p^a_{(i)} \, \} \, $ (with $ \,
i=1,2,...,N\, $) of the particle momenta are defined with respect to the frame  $ \, LS
\, $.  In addition to $ \, {\cal P}(t_0) \, $, the  $ \, (t = t_0) \, $ tessellation may
possibly contain several polygons; the complete description  of the initial
configuration includes all the lengths and  velocities of the polygon edges.   

During a  short time interval after $ \, t = t_0 \, $, one has a ``linear" evolution.
Because of the constant velocities of the edges, the length of each edge is a linear
function of time, the angles between edges and the particle momenta are constant in
time. When the length of one edge vanishes or when one of the corners hits an edge, one
has a transition in which the structure of the tessellation undergoes a local change.
So, the nontrivial part  of the evolution consists of a sequence of transitions; in each
transition  the shape and the number of polygons, together with the velocities of the
edges,  are in general modified according to a given set of rules [6]. 

After a finite time interval $ \, \Delta t = t_1 - t_0 \, $ in which no big-crunch
occurs, one finds a tessellation in which the particles are in general placed at the
corners of different polygons. In order to compare the initial $ \, (t=t_0 )\, $ and the
final $ \, (t=t_1 )\, $ values of the particle momenta, we need to refer these values to
a unique coordinate system; we shall choose the Lorentz frame $ \, LS \, $ as reference
system.  

Let $ \, Q_0 \, $ be a point in the interior of $ \, {\cal P}(t_0) \, $; we shall denote
by $ \, Q_1 \, $ the point in $ \, M \, $ which has the same spatial coordinates as 
$ \, Q_0 \, $ and has time coordinate $ \, t_1 \, $ with respect to the frame $ \, LS \,
$. We assume that $ \, Q_1 \, $ does not belong to the worldline of a particle; for, if
this is the case, we can simply modify the spatial position of  $ \, Q_0 \, $  so
that the corresponding $ \, Q_1 \, $ does not belong to a worldline. In a neighbourhood
$ \, \Omega \subset M \, $ of $ \, Q_1 \, $ the spacetime is flat and one can
introduce  local coordinates which are adapted to the Lorentz system $ \, LS \, $.  
The  piece in $ \, \Omega \, $  of the spatial surface $ \, t = t_1 \, $ can be extended
in $ \, M \, $  and one will eventually reach the worldlines of the particles.
This extension is not unique, in general. In order to remove all ambiguities, we
shall require that the extended  surface $ \, t = t_1 \, $ is star-shaped with respect to
the geodesic lines of minimal (spatial) distance emerging from $ \, Q_1 \, $; this simply
means that if a point $ \, Q \, $ belongs to the surface, then all the points of the
(minimal-length) geodesic connecting $ \, Q \, $ with $ \, Q_1 \, $ also belong to the
surface. This extension of the spatial surface $ \, t = t_1 \, $
determines a polygon $ \, {\cal P}(t_1) \, $ with the property [6,7] that each particle 
is placed at a corner of $ \, {\cal P}(t_1) \, $. Moreover,  one can introduce a
new $ \, (t=t_1) \, $ tessellation of space containing the tile $ \, {\cal P}(t_1) \, $.
With respect to the Lorentz frame $ \, LS \, $ of $ \, {\cal P}(t_1) \, $, which is the
same frame of $ \, {\cal P} (t_0) \, $, the final $ \, (t=t_1 )\, $ values of the
particle momenta  will be denoted by  $ \, \{ \, k^a_{(i)} \, \} \, $. 

To sum up, with respect to a fixed Lorentz system, the time evolution results in a
modification of the particle momenta 
$$
U(t_1 , t_0 ) \; : \; \{ \, p^a_{(i)} \, \}\; \longrightarrow \; \{ \, k^a_{(i)} \, \}
\qquad , 
$$
where the map $ \, U(t_1 , t_0 ) \, $ depends on the initial $ \, (t=t_0) \, $ data.
We shall conclude this section by recalling the general structure of $ \, U(t_1 , t_0 )
\, $. 

Consider a set  of oriented closed paths $ \, \{ \, \gamma_i \, \} \, $ on $ \,
{\cal P} (t_0) \, $ which are based on the point $ \, Q_0 \, $ and have no intersection
with the particle worldlines. Let the loop  $ \, \gamma_i \, $ be simply linked with the
worldline of the $ \, i$-th particle; the paths $ \, \{ \, \gamma_i \, \} \, $ are closed
because some of the edges of   $ \, {\cal P} (t_0) \, $ must be identified in the
tessellation.  The effects of a parallel transport of a vector
along the loop $ \, \gamma_i  \, $ can be described by means of a Lorentz transformation $
\, H(\gamma_i ) \, $ acting on the tangent space in $ \, Q_0 \, $. Since $ \, M \, $ is
flat in a neighbourhood of $ \, Q_0 \, $, the Lorentz frame $ \, LS \, $ canonically
defines a reference system in the tangent space at $ \, Q_0 \, $. 
The spacetime curvature is concentrated on the particle worldlines, so $ \, H(\gamma_i )
\, $ is invariant under smooth deformations of $ \, \gamma_i \, $. Moreover, one has $ \,
H(\gamma_i^{-1} ) = H^{-1}(\gamma_i )\, $ where  $ \, \gamma_i^{-1}\, $ represents $\,
\gamma_i \, $  with reversed orientation. Now, with a suitable choice of the loops 
$ \, \{ \, \gamma_i \, \} \, $ and a given (clockwise) prescription for the loop
orientations, one finds [6] 
$$
H(\gamma_i ) \; = \; \exp \left ( \, 8 \pi G \, p^a_{(i)} J_a \, \right ) \; = \; \exp
\left ( \, {\widehat p_{(i)}} \, \right ) \quad , 
$$
where $\, \{ \, J_a \, \} \, $ are the generators of the Lorentz group and $ \, \{
\, p^a_{(i)} \, \} \, $ are the components of the momentum of the $i$-th particle with
respect to the reference system  $ \, LS \, $. Since all the particles are placed at 
the corners of the polygon $ \, {\cal P}(t_0) \, $ whose boundary -without edges
identifications- has the topology of  $ \, S^1 \, $,  an ordering of the particles turns
out to be fixed modulo cyclic permutations; let us assume that  the natural ordering
of the index $ \, i = 1,2,...,N \, $ corresponds to consecutive particles on the boundary
of $ \, {\cal P}(t_0) \, $.  The initial $ \, (t = t_0) \, $ particle data can conveniently
be expressed by means of an ordered set $ \, I \, $ of $\, SO(2,1) \, $ group elements  
$$
I \; = \; \{ \, H(\gamma_1) \, , \, H(\gamma_2) \, , \dots , \, H(\gamma_N) \,\} 
\; = \; \{ \, e^{{\widehat p}_{(1)}} \, , \,
e^{{\widehat p}_{(2)}} \, , \dots , \, e^{{\widehat p}_{(N)}} \, \} \qquad .  
$$
We now need to introduce the Lancaster-Sasakura representation [8]  of the braid group $
\, B_N \, $ acting on $ \, I \, $; the generator $\, \sigma_j \, $ of $ \, B_N\, $ (with $
\, j=1,2,...,N-1 \, $) is represented by 
$$
\sigma_j \; : \; \{ \, e^{{\widehat p}_{(1)}} \, , \dots , \, e^{{\widehat p}_{(N)}} \, \}
\; \mapsto \; \{ \, e^{{\widehat p}_{(1)}} \, , \dots , \, e^{{\widehat p}_{(j-1)}} \, ,
\, e^{{\widehat p}_{(j+1)}}\, , \, e^{-{\widehat p}_{(j+1)}}e^{{\widehat
p}_{(j)}}e^{{\widehat p}_{(j+1)}}\, , \cdots , \, e^{{\widehat p}_{(N)}} \, \} \; . 
$$
Let us denote by $ \, F \, $ the ordered set of $\, SO(2,1) \, $ group elements 
$$
F \; = \;  \{ \, e^{{\widehat k}_{(1)}} \, , \,
e^{{\widehat k}_{(2)}} \, , \dots , \, e^{{\widehat k}_{(N)}} \, \} \qquad ,   
$$
where $ \, \{ \, k^a_{(i)} \, \} \, $ are the final $ \, (t = t_1) \, $ particle momenta. 
One can show [6,-,9] that the time-evolution map is  given by an element  $ \,
g(t_1, t_0) \in B_N \, $ in the Lancaster-Sasakura representation  
$$
U(t_1 , t_0 ) \; : \; I \;  \rightarrow \; F \; = \; g(t_1, t_0) \, I \qquad . 
\eqno(2.1)
$$
The particular element $ \, g(t_1, t_0) \in B_N \, $ which enters equation (2.1) depends
on the initial positions and velocities of the particles. Any element of the braid
group can be written as an ordered product of generators. So the time evolution can be
interpreted as the result of a certain number of consecutive elementary processes; each
elementary process is associated with a generator of the braid group. 

\bigskip 

\noindent {\bf 3. Computation rules.} ~For any specific choice of the
initial lengths and edge velocities of the starting tessellation, 't~Hooft rules determine
a unique time evolution.  With a large number of particles, the complete evaluation of the
exact microscopic state of the system is rather laborious [6]. On the other hand, we are
interested in the thermodynamic  properties of a large number $ \, N \, $ of
gravitating particles. Therefore we shall introduce  certain approximations which simplify
the computation of the macroscopic variables. The main idea consists of adopting a
iterative method -based on some  random process- to determine the Braid group element  $\,
g(t_1, t_0) \, $ which specifies the evolution (2.1) of the particle momenta. 

If one has a 3-dimensional classical gas in standard conditions of temperature and
pressure, one can approximate the molecular motion by a sequence of diffusion processes
among randomly chosen pairs of molecules. This is not an accurate approximation at the
microscopic level but turns out to be a  rather good method to evaluate macroscopic
variables. Similarly, for the 2-dimensional gravitational gas we will approximate the
time evolution by a sequence of ``scattering processes" involving randomly chosen pairs of
particles.  

Suppose that $ \, a \, $ and $ \, b \, $ are two integers which have been randomly chosen
among the set $ \, \{ \, 1,2,..., N \, \} \, $ with the condition $ \, a \not= b \, $.  
Suppose that,  for instance,  $ \, a < b \, $.   One can then define a particular element
$ \, h(a,b) \, $ of the braid group $ \, B_N \, $  which is the product of  $\, (b-a) \,
$ generators 
$$
h(a,b) \; = \; \sigma_{b-1} \cdots   \sigma_{a+1} \cdot \sigma_a \qquad . 
$$
The elementary process associated with the integers $ \, a \, $ and $ \, b \, $ can be
interpreted as a scattering process involving the $\, a$-th and $ \, b$-th particles.
After this scattering, the particle momenta are described by the new ordered set of
$\, SO(2,1) \, $ elements 
$$
I^\prime \; = \;  h(a,b) \, I \qquad , 
\eqno(3.1)
$$
or  
$$
I^\prime \; = \;  h^{-1}(a,b) \, I \qquad . 
\eqno(3.2)
$$
Possibilities (3.1) and (3.2) are assumed to have  equal probabilities to occur. After
this first step, one has simply to repeat the procedure: one determines two new random
numbers $ \, a^\prime \, $ and $ \, b^\prime \, $ and applies $ \, h(a^\prime ,b^\prime )
\, $ (or $ \, h^{-1}(a^\prime ,b^\prime ) \, $, with probability 1/2) to the ordered
set  $ \, I^
\prime \, $ in order to find $ \, I^{\prime \prime} \, $, and so on. 

One can use a simpler recipe to approximate the time evolution. At each step of this
new recipe, one randomly determines an integer $ \, a \in \{ 1,2,...,N-1\} \, $ and the
modification of the particle momenta is described, with equal probabilities, by the
generator  $ \, \sigma_a \, $ or its inverse $ \, \sigma^{-1}_a \, $ in the
Lancaster-Sasakura representation.  We have verified that, in all the considered
examples,  both methods lead to the same conclusions. 

Our computation rules only concern the values of the particle momenta; the spatial
positions of the particles and their mutual distances have been ignored.  In the ordinary
3-dimensional case of a gas inside a finite volume, one can safely ignore the
spatial positions of the molecules because the distribution of equilibrium is
in fact homogeneous. So, as long as the gas is approaching the thermal equilibrium, the
actual position of each molecule plays no role  in the evaluation of the
macroscopic variables.  For this reason, the assumption of a homogeneous
equilibrium distribution could be introduced to justify our computation rules. 
Actually,  the 2-dimensional gravitating gas that we consider is not evolving inside a
container with fixed walls; consequently, the equilibrium particle distribution -when it
exists- is not necessarily homogeneous in space.  Yet, the computation rules that
we use to approximate the evolution of the particle momenta  are expected to be valid
anyway  because each elementary process involving  two gravitating
particles in (2+1) dimensions does not depend on the distance between these particles. 

\bigskip 

\noindent {\bf 4. Conservation law.} ~Since we ignore the spatial positions of the
particles as well as the whole ``extension" of the system, we can only compute how the
thermodynamic potentials depend, for instance, on the internal energy $ \, U \, $ of the
system.  So we need to give a definition of  $ \, U \, $ and discuss the associated   
conservation law. 

The components $ \,  p^a = (p^0, \vec p \, )  \, $ of the momentum of a classical pointlike
particle of mass $ \, m \, $ can be determined by measuring the velocity of the particle
with respect to a local Lorentz frame. Because of  Lorentz covariance, the energy $ \,
p^0 \, $ and the spatial components $ \, \vec p \, $ of the momentum have the usual
dependence on the mass and on the velocity  of the particle. One has $ \, (p)^2 = p^a p_a
= m^2 \, $. For a system of $ \, N >1 \, $ gravitating particles, the sum of the energies
of the particles is not a conserved quantity.  In the presence of gravitational
interactions, the ordinary conservation law for the 3-momentum gets modified; the
new conservation law [7] for the 3-momentum is Lorentz covariant and can be expressed in
terms of  $\, SO(2,1) \, $ group elements. Since the time evolution of the particle
momenta is given in  equation (2.1), it is easy to verify that, whatever 
$ \, g(t_1, t_0) \in B_N \, $ might be,  the composite element 
$$
e^{{\widehat p}_{(1)}} \cdot
e^{{\widehat p}_{(2)}} \cdots e^{{\widehat p}_{(N)}}   \; = \;  
e^{{\widehat p}_{(tot)}} \; = \; 
\exp \left ( \, 8 \pi G \, p^a_{(tot)} J_a \, \right ) 
\eqno(4.1)
$$
is conserved;  $ \, \{ \,  p^a_{(tot)} \, \} \, $ are the components of the total
3-momentum  of the $ \, N \, $ particles.  In (2+1) dimensions the gravitational field does
not describe propagating degrees of freedom; indeed, the total 3-momentum only depends on
the variables associated with the particles. No Newtonian or potential energy is present in
(2+1) dimensions, and in fact $ \, \{ \,  p^a_{(tot)}  \, \} \, $ do not depend on the
positions of the particles. As expected, in the $ \, G \rightarrow 0 \, $ limit,
expression (4.1) reproduces the usual additive conservation law of 3-momentum which is
valid in ordinary Minkowski space.  

The total 3-momentum admits an intrinsic definition. Consider an oriented loop $ \, \gamma
\, $ in $ \, M \, $ based on the fixed point $ \, Q_0 \in M \, $ and let $ \, \Sigma \, $ be an 
oriented disc in $ \, M \, $ whose boundary coincides with $ \, \gamma \, $. Each
particle worldline has a natural orientation and its intersection points with $ \, \Sigma
\, $ have definite signatures.  We will say that a given worldline is enclosed by $ \,
\gamma \, $ when the algebraic sum of its intersections with $ \, \Sigma \, $ is equal to
unity. Suppose now that all the worldlines of the particles are enclosed by 
$ \gamma \, $; the path-ordered  integral of the spin connection (which must be defined
in a smooth coordinate system) along $ \, \gamma \, $ determines the total
3-momentum. In facts, the $ \, SO(2,1) \, $ element which describes the effects of a
parallel transport of vectors along $ \,  \gamma \, $  is given by 
$$
H(\gamma ) \; = \; \exp \left ( \, 8 \pi G \, p^a_{(tot)} J_a \, \right )  \qquad . 
\eqno(4.2) 
$$
When the total 3-momentum is timelike, one can find a Lorentz transformation $ \, \Lambda
\, $ such that  
$$
\Lambda \cdot \exp \left ( \, 8 \pi G \, p^a_{(tot)} J_a \, \right ) \cdot \Lambda^{-1} \;
= \; \exp \left ( \, 8 \pi G \, U J_0 \, \right ) \qquad ,   
\eqno(4.3)
$$
where $ \, U \, $ is called the internal energy of the particles system. The
generator of rotations $ \, J_0 \, $ has integer eigenvalues and then  $ \, 8 \pi G U \, $
is apparently defined only modulo $ \, 2 \pi \, $. Actually this is not the case;  the
value 
$ \, U \, $ of the internal energy is well defined in the entire real axis. Indeed, 
consider a given parametrization of $ \, \gamma \, $ with variable $ \, \tau
\in [0,1] \, $; since the spin connection takes values in the algebra of the Lorentz
group,  the path-ordered integral of the spin connection in the interval  $ \,
[0, \tau ] \, $ defines a continuous map  $ \, \psi : \tau \mapsto \psi_\tau \in \widetilde
{SO}(2,1) \, $  where $ \, \widetilde {SO}(2,1) \, $ is the infinite cyclic covering of $
\, SO(2,1) \, $ [10].  By construction, one has $\, \psi_0 =$ identity and the projection
of  $ \, \psi_1 \, $ into $ \, SO(2,1) \, $ coincides with $ \, H(\gamma ) \, $. 
The fact that the value of $ \, U \, $ is well defined in the whole real axis should not be
surprising;   something similar occurs with the time defined by a watch. Even if the
position of the minute-hand is specified by an angle which is  defined modulo $ \, 2
\pi \, $, there is no ambiguity at all in determining, by continuously observing the watch,
the time interval elapsed with respect to a given moment.  

\bigskip 

\noindent {\bf 5. Numerical results.} ~In order to complete the description of the
computation rules, we need to discuss the specification of the initial data which consists
of three steps:   

\begin{enumerate} 

\item  it is assumed that all the particles have the same mass $\,m\,$;
  a set of values $\,\{ \, q^a_{(i)} \, \}\,$ for the particle momenta is 
  fixed by choosing the spatial components $\,\{ \, \vec q_{(i)} \, \}\,$ randomly within
  a finite domain; 

\item for a given set of values $\,\{ \, q^a_{(i)} \, \}\,$ determined in point~1, the
  corresponding total 3-momentum is computed by means of the relation (4.1);  

\item if the total momentum is time-like, by means of a Lorentz transformation $\,\Lambda\,$ the
particles system is
  taken to its rest frame.  $\,\Lambda\,$ is fixed according to equation (4.3) 
  which also determines the internal energy $\,U\,$ of the system. The set of 
  particle momenta with respect to the system rest frame $\,\{ \, p^a_{(i)} = {\Lambda^a}_b
    q^b_{(i)} \, \}\,$ represents the ordered set of initial particle momenta on which the
  time evolution acts. 

\end{enumerate}

In the various numerical simulations that we have performed,  the number of particles has been
chosen within  $\,10^2 \leq N \leq10^3\,$ and the values of the masses belong to the range $\,0
\leq m \leq 10^{-2}/ 4G\,$. Since the total 3-momentum has been determined by means of equation
(4.1), the values of the mass and of the initial momentum of each particle have been chosen to be
``small enough'' so that no $\, 1/4 G\,$ ambiguity resulted in the computation of the total energy.
Instead of performing the Lorentz transformation mentioned in point 3,
in several occasions we have used a different method to bring the system in its rest
frame; namely, to the set of the initial $\,N\,$ particles we have simply added a new
particle with a suitably chosen momentum so that the resulting spatial components
$\,\vec p_{(tot)}\,$ vanish. This last method is convenient because it avoids strong
correlations among initial particle momenta. 

\bigskip

\noindent {\bf 5.1. Open universe.} ~We have done simulations consisting of approximatively
$\,4\times 10^9\,$ random interactions and we have verified that, as long as $\,U < 1/4 G\,$, the
simulations always reached a stationary state within the first 10\% of the interactions, which we
excluded from the final data. The numerical results are expressed by means of the equilibrium
distribution function $\,f(\vec p\,)\,$ such that
\[
  dN \; = \; f(\vec p\,) \; d^2p
\]
represents the particle density in momentum space. The normalization condition is 
\[
  \int d^2p \; f (\vec p\,) \; = \; N \quad . 
\]
Since $\,f(\vec p\,)\,$ is isotropic, it only depends on  $\,|\vec p\,|\,$; so it is
convenient to introduce a distribution function $ \, \rho(p^0)\, $ for the energy $\,p^0=\sqrt{m^2 +
|\vec p\,|^2}\,$ of a single particle:
\[
  \rho(p^0) \; = \; \frac{2\pi \, p^0}{N}  \, f(|\vec p \,|) \; = \; \frac{2\pi \, p^0}{N}  \,
f(\sqrt { (p^0)^2 - m^2 }) \quad . 
\]
The  $\,\rho\,$ normalization is 
\[
  \int_m^{+\infty} dp^0 \; \rho(p^0) \; = \; 1 \quad . 
\]

\begin{figure}[htb!]
  \begin{center}
    \begin{picture}(0,0)(0,0)
    \end{picture}
    \put(-30,120){$\rho(p^0)$}
    \includegraphics[width=12cm,height=8.2cm]{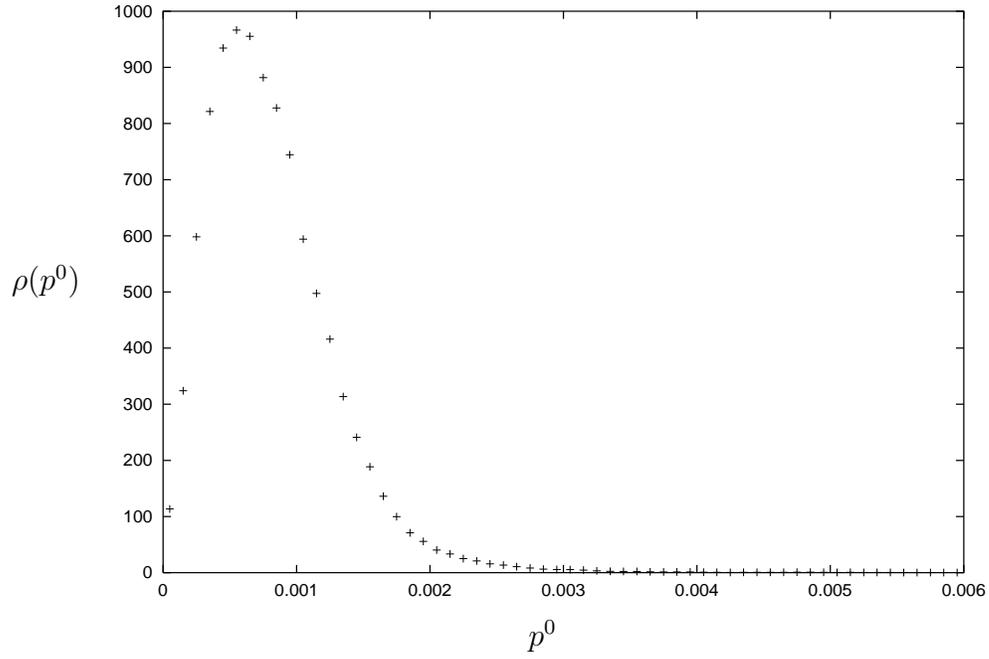}
    \\ $p^0$
    \caption{plot of $\,\rho(p^0)\,$ at $N=1000,\ m=0,\ 4GU=1.296\times 10^{-1}\,$}
    \label{1-1}
  \end{center}
\end{figure}

\begin{figure}[htb!]
  \begin{center}
    \begin{picture}(0,0)(0,0)
    \end{picture}
    \put(-30,120){$\rho(p^0)$}
    \includegraphics[width=12cm,height=8.2cm]{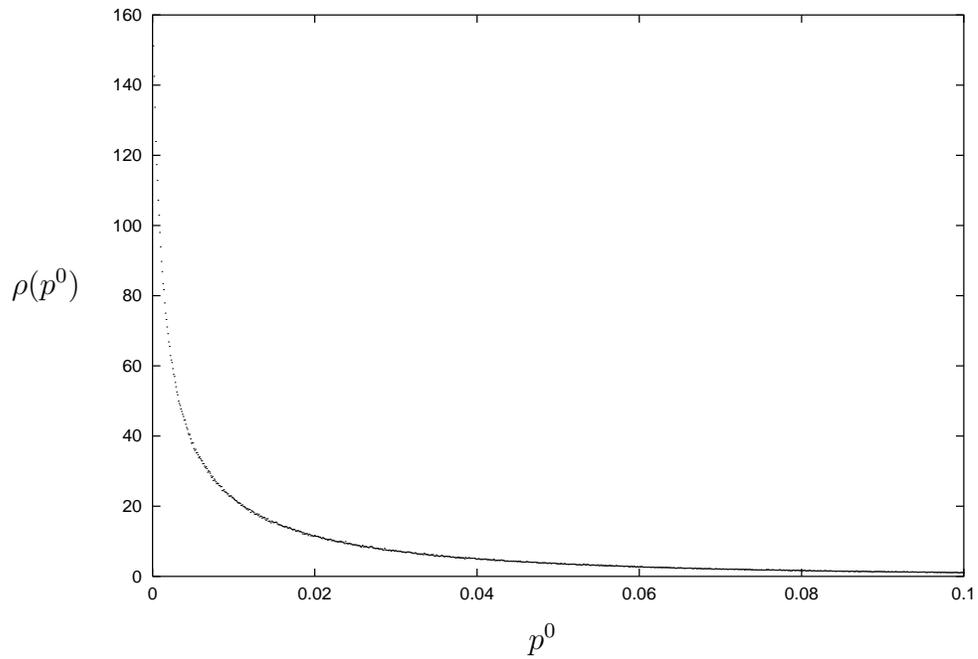}
    \\ $p^0$
    \caption{plot of $\,\rho(p^0)\,$ at $\,N=1000,\ m=0,\ 4GU=9.9968 \times 10^{-1}\,$}
    \label{1-2}
  \end{center}
\end{figure}

\bigskip 
When $\,U < 1/4G\,$, we observed that the system always
reaches a stationary state. Two examples of computed distribution function  $\,\rho\,$ 
are shown in Figure \ref{1-1} and \ref{1-2}. In both cases, $\,N=10^3\,$ and $\,m=0\,$.
Figure \ref{1-1} refers to the case in which $\,4GU= 1.296\times 10^{-1}\,$, whereas in Figure
\ref{1-2} one has $\,4GU= 9.9968\times 10^{-1}\,$. In all the figures,  energies and momenta are
expressed in units of $\,1/8\pi{}G\,$. The qualitative behaviour of the gravitating gas does not
depend on the value of $ \, m \, $ provided $ \, N m < 1/4G \, $. 

In the low energy limit $\,U \ll 1/4G\,$  the equilibrium distribution $\,f(\vec p\,)\,$ is well
fitted by the Boltzmann distribution that we shall derive later. This is in agreement with the fact 
that, for $\,G\rightarrow0\,$, the energy conservation law reduces to the classical one. 

As $\,U\,$ grows, $\,f(\vec p\,)\,$  differs from the Boltzmann distribution substantially.
The high-energy tail of $\,\rho (p^0)\,$ is rather well fitted by the function 
$\,a/(p^0+b)\,$ with a truncation at a certain $\,p^0 = \ov{p}^0(U) \,$. 

\begin{figure}[htb!]
  \begin{center}
    \begin{picture}(0,0)(0,0)
    \end{picture}
    \put(-30,120){$ \bra \, p^2 \, \ket $}
    \includegraphics[width=12cm,height=8.2cm]{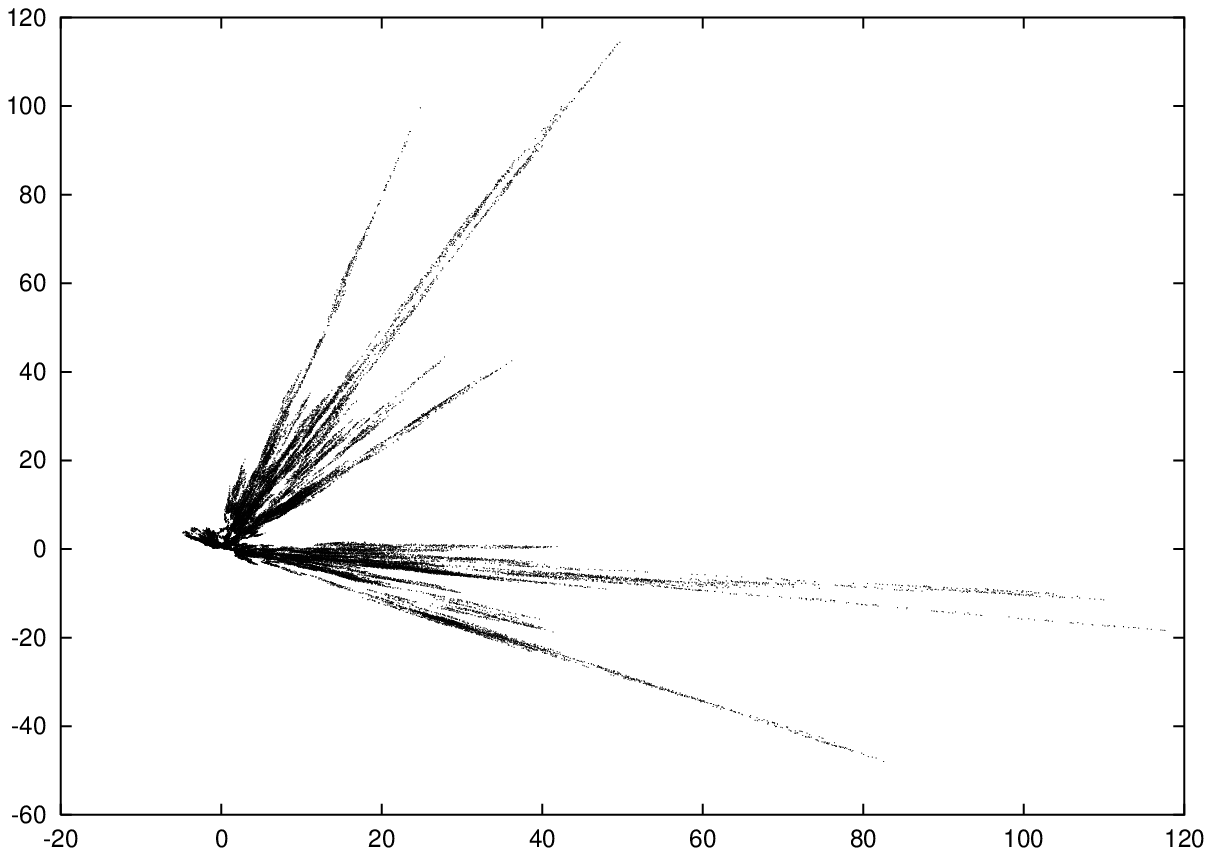}
    \\ $\bra \, p^1 \, \ket $
    \caption{plot of $\, \bra \, \vec p\; \ket \, $ at $\,N=100,\ m=7.161\times10^{-3},\ 4GU=9.9980
\times 10^{-1}\,$}
    \label{1-jet}
  \end{center}
\end{figure}

A new phenomenon happens when the value of $\,U\,$ approaches $\,1/4G\,$: the whole 
gas strongly correlates, all the particle momenta tend to become parallel and the overall
direction fluctuates randomly as the system evolves. Figure \ref{1-jet} shows
the mean value $\, \bra \, \vec p\; \ket \, $ of $\,\vec p\,$ in a finite time interval for
$\,4GU\,$ almost equal to unity.  The time evolution presents large fluctuations in $\,|\vec p
\,|\,$ which appear as spikes in the picture.

\newpage

\noindent {\bf 5.2. Spacelike total momentum.} ~In this case, numerical simulations show that the
system is unstable and no equilibrium distribution exists. During the evolution, the energies of the
particles tend to increase overcoming whatever fixed value. The way energies diverge is
quite peculiar: at first, all energies tend to increase slowly and, after a while, they grow
faster and faster.  The higher the energy is, the quicker it increases (positive feedback) so nothing
can prevent energies to diverge.  At this stage, Gott pairs are created and a ``jet'' appears.
Two consecutive particles form a Gott pair [11] when the non-Abelian composition $ \, p \, $ 
of their momenta, 
$$
\exp  \left ( \, {\widehat p}_{(1)} \, \right ) \cdot
\exp  \left ( \, {\widehat p}_{(2)} \, \right ) \; = \;
\exp  \left ( \, {\widehat p} \> \right ) \quad , 
$$
is space-like $ \, (p)^2 = p^a p_a < 0 \, $. A jet is a set of a few consecutive particles -usually
less than 10- whose momenta are almost parallel and whose energies are much bigger than the
gas average. Jets have a remarkable property: when particles in the same jet interact, their 
momentum is in general strongly boosted in the jet direction and their energies can increase
even by $\,10\div30\,$ orders of magnitude in a single transition.  Note that, as demonstrated in
[12,13], Gott pairs cannot be created when $ \, 4GU < 1 \, $.  

Simple qualitative considerations show that a universe with spacelike total
momentum is either exploding or imploding. Results obtained from our simulations of
such universes always correspond to the imploding case, where particles looping
around the whole universe always get boosted in the same direction as their velocity. 
In an exploding universe, particles with speed lower than the global expansion rate cannot
complete a loop around the universe but, by construction, our algorithm does not take into account
this global effect. The instability that we find for spatial total momentum is in agreement with the
big-crunch phenomena studied in [6]. 

\bigskip

\noindent {\bf 5.3. High energies.} ~For a spacetime of topological type $ \, S^2 \times \erreuno
$, the Gauss-Bonnet equality implies $ \, 4GU = 2\, $. Suppose that one or more particles are very
far from all the others and do not interact with them; these particles can be neglected so that it
makes sense to consider a particle system with energy  $\,1 \leq 4GU < 2\,$. The particular case $
\, 4GU= 1\, $ also corresponds to a (globally static) cylindrical universe. 

In the whole range $ \,1 \leq 4GU \leq 2\,$ we observed the same behaviour as for spacelike total
momentum: Gott pairs, jet formation and  instability due to divergent energies. This suggests that,
in our approximation, a closed universe usually develops Gott pairs.

Because of the finite precision of numerical simulations, it is impossible to constraint
$\,U\,$ to be exactly equal to a predefined value $\,X\,$. Using double precision floating point
numbers (8 bytes) we can only obtain simulations with $\,|U-X|\lesssim 10^{-15}\,$ and $\,U\,$
fluctuates in that range as the simulation proceeds. We have even seen, though very
rarely, simulations with $\,4GU<1\,$ explode due to $\,4GU\,$ becoming $\,>1\,$ by
finite precision errors.
For this reason, it is not possible to distinguish the gas behaviour at
$\,4GU=1\,$ (or $\,4GU=2\,$) from the behaviour at very near values of $\,U\,$.
What we have actually seen is a strong instability in both cases.

\bigskip 
\noindent {\bf 6. Classical statistics.} ~In the next section we shall compute the entropy of the
gravitating gas when the energy satisfies $\, 4GU<1\,$ and we shall compare the outcome of the
simulations with the entropy of a non-gravitating relativistic Boltzmann gas in (2+1)
dimensions. Here we recall a few classical statistical mechanics results.

For a classical relativistic gas with 2-dimensional volume $ \, V \, $, 
the Maxwell-Boltzmann's distribution $\,f_0\,$ takes the form
\[
  f_0 (\vec p\,) \; = \; \left ( {N\over V} \right ) {e^{( \, m / kT )}\over 2 \, \pi \, k \, T \, (
\, m + kT \, )} \; \,  \exp \left ( \, - \, {\sqrt { (\vec p \, )^2 \, + \, m^2 } \over kT} \, \right
)
\]
and has normalization 
\[
  \int d^2x \, d^2p  \; f_0 (\vec p\,) \; = \; N \quad.
\]
The pressure $\,P\,$ can be computed in the usual microscopic way: one has to sum the variation of
momenta over the particles hitting a unitary surface.  The result is the same as for a perfect gas
\[
  P V \; = \; N k T \quad . 
\]
The internal energy turns out to be
\[
U  \; = \; U ( T , V) \; = \; N \, \bra \, p^0  \, \ket \; = \; N kT \left( \frac{m + 2kT}{m + kT}
\right)  \, + \, N m  \quad .
\]
According to Boltzmann's $\,H$-theorem, the entropy is
\[
  S \; \equiv \; -k  \int d^2x \, d^2p \, f_0 \log(f_0) \; =\; kN \left(
  \frac{m + 2kT}{m + kT} - \log \frac{N}{V} + \log \Big( kT (m + kT) \Big)
  \right)
\]
which satisfies the fundamental relation
$$
  \frac{\de S(U,V)}{\de U} \; = \; \frac{1}{T} \quad . 
\eqno(6.1) 
$$
When $\,m =0\,$ one obtains
$$
  U \; = \; U(T,V) \; = \;  2 N kT \eqno (6.2)
$$
$$
  S \; = \; S ( U , V) \;= \; kN \left( 2 - \log \frac{N}{V} + 2 \log {U\over 2N} \right) 
  \quad . \eqno (6.3)
$$
Note that equation (6.2) is in agreement with the Equipartition Theorem because the dispersion
relation takes the form $ \, \sqrt { | \vec p \, |^2 }\, $. 

\bigskip
\noindent {\bf 7. Statistical properties of (2+1) dimensional gravitating gas.} ~Numerical simulations
showed that the open universe with energy $\, 4GU<1\,$ is stable, i.e. the system always reaches a
stationary state, therefore we shall now examine the statistical properties of such a gas. In
particular, we will study the entropy and the temperature of the system.

In order to verify that the stationary state corresponds to thermal equilibrium,
we have checked that the connected part of the 2-particle probability distribution 
\[
  f_{2c}(p_i,p_j) = f_2(p_i,p_j) - f(p_i) \cdot f(p_j)   
\]
is numerically negligible. We again apply Boltzmann's $\,H$-theorem to give a 
numerically usable definition of entropy; considering only particle momenta, one has  
$$
 S \; \equiv \;  - k  \int d^2p\; f(p) \, \log(f(p))
 \quad . \eqno (7.2)
$$
For the plots, a ``reduced'' definition of $\,S\,$ has been used:
\[
  \ov{S}
  \; = \; \frac{S}{Nk} - \log\left(\frac{2\pi}{N}\right)
  \; = \; - \int_m^{+\infty} dp^0 \rho(p^0) \log\left(\frac{\rho(p^0)}{p^0}\right) \quad .
\]

\begin{figure}[htb!]
  \begin{center}
    \begin{picture}(0,0)(0,0)
    \end{picture}
    \put(-30,120){$\ov{S}$}
    \includegraphics[width=12cm,height=8.2cm]{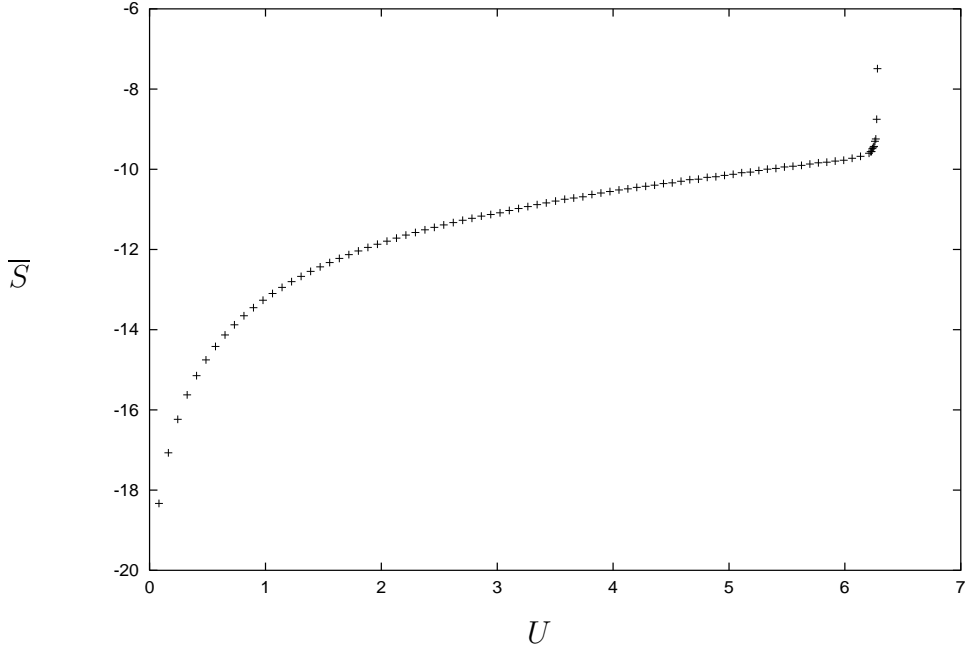}
    \\ $\,U\,$
    \caption{plot of $\,\ov{S}\,$ at $\,N=1000,\ m=0\,$}
    \label{1-3}
  \end{center}
\end{figure}

As one can deduce from Figure~4, the values of $\,S\,$ that we found are very well fitted by
$\,S\propto\log(U) + {\rm const}\,$ for energies up to $\,4GU\lesssim0.8\,$.  As $\,U\,$ grows,
$\,\de S / \de U\,$ decreases more slowly than expected. At a certain value $\,U=\ov{U}\,$, $\,\de S
/ \de U\,$ stops decreasing and
$\,S\,$ has a flex $\,(\desq S / \de U^2=0)\,$.
For $\,U>\ov{U}\,$, the second derivative of $\,S\,$ is $\,>0\,$ and increases as
$\,4GU\rightarrow1\,$. In the same limit, the entropy $\,S\,$ seems to diverge.
This behaviour is supported by the following argument.
For a given $\,U\,$, let $\,f_m\,$ be the maximum value of $\,f(|\vec p\,|)\,$.
At $\,4GU>1\,$ we already said that simulations diverge, 
and then it is reasonable to assume that
$$
 \lim_{4GU\rightarrow1}f_m=0 \quad . \eqno (7.3)
$$
From the definition (7.2) of S we get
\[
  S \; = \; -k \int d^2p \; f(p) \, \log(f(p)) \; > \; -k \, \log(f_m) \int d^2p \; f(p) \; = \;
 -Nk \, \, \log(f_m) \; , 
\]
then by means of (7.3) one finds
\[
  \lim_{4GU\rightarrow1}S(U) \; = \; +\infty \quad .
\]

\noindent Finally, for $\,4GU>1\,$ there is no equilibrium state, so $\,S\,$ is not defined in
that range.

\bigskip

\noindent {\bf 7.1 Temperature} ~The standard interpretation of the entropy behaviour as a 
function of $\,U\,$ uses the relations
$$
  \frac{1}{T} \; = \; \frac{\de S(U,V)}{\de U}
\eqno(7.4)
$$
and
$$
  \frac{1}{C} \; = \; - T^2 \frac{\desq S(U,V)}{\de U^2}
  \quad . 
\eqno (7.5)
$$
According to these equations, the temperature $\,T\,$ has a maximum at $\,U=\ov{U}\,$ and the heat
capacity $\,C\,$ diverges at $\,U=\ov{U}\,$ and is negative at $\,U>\ov{U}\,$.
Such results look rather strange  as they  violate the thermodynamical inequalities. 
In a standard statistical system with additive $ \, S \, $ and $ \, C \, $, the violation of these
inequalities implies instability. This is not true in our system; when $\,4GU <1\,$ the gravitating
gas reaches the equilibrium. The standard instability theorems do not apply to our system because 
$\,S\,$ and $ \, C \, $  are not additive. 

In order to illustrate the peculiar properties of the
gravitating gas, consider for example two identical systems, each one in equilibrium, which are
isolated.  When these two systems  are put in thermal contact, in general the resulting system is
not in equilibrium.

The temperature $ \, T \, $  defined in equation (7.4) is not a convenient variable to deduce the
stability properties of the gravitating system.  For the Boltzmann gas, one can give different
definitions of temperature that are all equivalent. In the case of a gravitating gas, these
definitions are no longer equivalent. Firstly, we have two energies that characterize the
equilibrium state:  $ \, N {\cal E} \; = \; N \bra \, p^0 \, \ket \, $ and $ \, U \, $. 
Due to the peculiar energy conservation law in (2+1) gravity, $\, N \cal E \,
$ and $\, U \, $ are  not trivially related. Secondly, we shall introduce the temperature in two
ways: by means of appropriate Equipartition Theorem (6.2) or by means of the entropy (6.1). 
For simplicity we shall concentrate on the $\,m=0\,$ case. By taking into account all the
possibilities, the various definitions of temperature are:
$$
  T_1 \; = \; \frac{\cal E}{2k} \qquad , \qquad T_2 \; = \; \frac{U}{2Nk}
$$
$$
  T_3 \; = \; N \left( \frac{\de S}{\de \cal E} \right)^{-1} \qquad , \qquad 
  T_4 \; = \; \left( \frac{\de S}{\de U} \right)^{-1} \quad . 
$$
$T_4\, $ coincides with definition (7.4).

\begin{figure}[htb!]
  \begin{center}
    \begin{picture}(0,0)(0,0)
    \end{picture}
    \put(-40,120){$kT$}
    \put(340,150){$\ds kN \! \left( \frac{\de S}{\de \cal E} \right)^{\! \! -1}$}
    \put(340,100){${\cal E}/ 2$}
    \put(340,65){$U/ 2N$}
    \put(340,15){$\ds k\left( \frac{\de S}{\de U} \right)^{\! \! -1}$}
    \includegraphics[width=12cm,height=8.2cm]{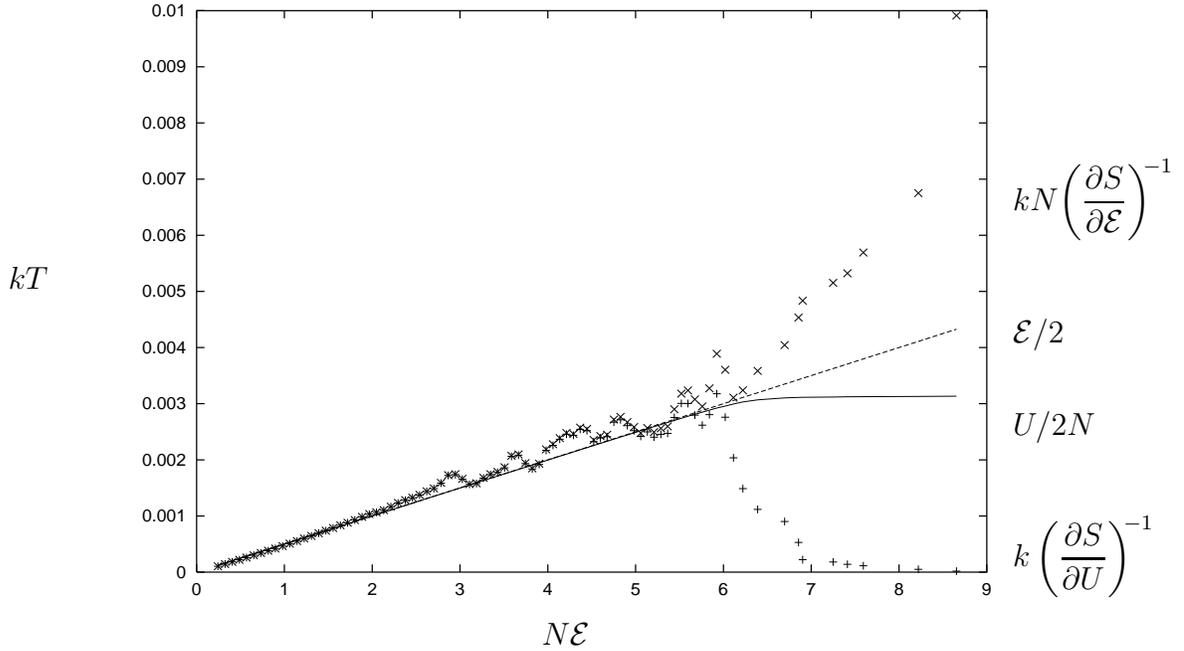}
    \\ $\,N{\cal E}\,$
    \caption{plot of $kT\,$ at $\,N=1000,\ m=0\,$}
    \label{1-5}
  \end{center}
\end{figure}

As we can see from the plot, all the definitions of $\,T\,$ are in good agreement with each other
as long as $\, 4GU\lesssim0.8 \,$, while at higher energies each
definition of $\,T\,$ has a different behaviour: when plotted as a function of $\,{\cal E}\,$,
$\,T_3 = N ( \de S/ \de {\cal E} )^{-1}\,$ grows more than linearly, while
$\,T_4 =( \de S/ \de U )^{-1}\,$ has the maximum we already anticipated, then drops quite
quickly. The other two ones $\,T_1 = {\cal E} / 2k\,$ and $\,T_2=U/2Nk\,$ have quite obvious
behaviours: the first one is linear, while the second asymptotically approaches $\, 1/8GNk\,$ as
$\,4GU\rightarrow1\,$.

By plotting $ \, S \, $ as function of $ \, \log {\cal E} \, $ (see Figure 6), one
notices that the fit $ \, S \propto \log {\cal E} \, $ is quite reasonable at all energies. This
suggests that
$
\, T_3 \, $ is in some sense the most suitable definition of temperature.  With respect to $ \, T =
T_3 \, $, the $\,4GU\rightarrow1\,$ limit corresponds to $ \, {\cal E} \to + \infty \, $ and
simultaneously $ \, T \to  + \infty \, $; moreover, the heat capacity is always finite and
positive. 

\begin{figure}[htb!]
  \begin{center}
    \begin{picture}(0,0)(0,0)
    \end{picture}
    \put(-30,120){$\ov{S}$}
    \includegraphics[width=12cm,height=8.2cm]{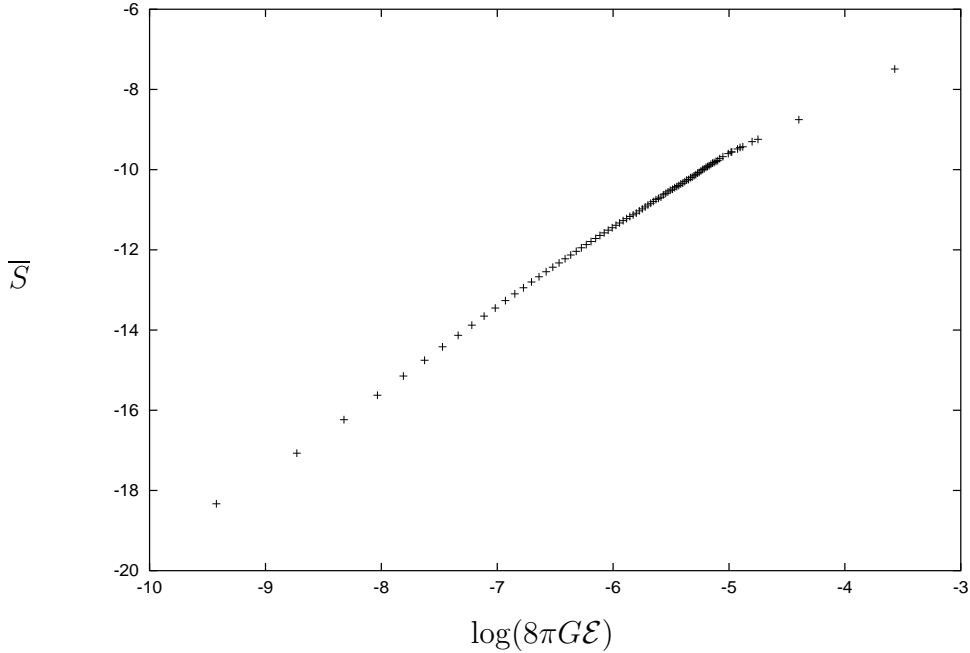}
    \\ $\, \log(8\pi{}G{\cal E})$
    \caption{plot of $\,\ov{S}\,$ at $\,N=1000,\ m=0\,$}
    \label{1-4}
  \end{center}
\end{figure}

\bigskip
\noindent {\bf 8. Conclusions.} ~A system of gravitating particles in (2+1) dimensions has peculiar
thermodynamic properties due to the non-Abelian structure of the energy-momentum conservation law. 
Depending on the total 3-momentum, numerical simulations show different system behaviours. 
When the internal energy $\, U \, $ satisfies $ \, 4 G U < 1 \, $, an equilibrium state is
reached; on the other hand, for space-like total momentum or when $ \, 4 G U >  1 \, $,
dynamical instabilities appear.  This is consistent with the interpretation of particle diverging
energies as a big-crunch phenomenon. 

In the presence of thermal equilibrium, we started with the distribution function in momentum
space and studied the thermodynamic properties of the gravitating particles. In the low energy
limit, which is equivalent to the low gravity limit, the system behaves as a Boltzmann relativistic
gas.  For $ \, 4 G U \to 1 \, $,  the entropy and the mean value of the kinetic energy  seem to
diverge.  

Starting from the various equivalent definitions of temperature in Boltzmann gas, we have
considered their behaviour also in the energy range of large gravity effects. These temperatures
turn out to be different.  One of them is well behaved also in the $\,4GU\rightarrow1\,$ limit
and does not violate the thermodynamic inequalities; in particular, the associated heat capacity is
always finite and positive. 

\vfill\eject

\vfill\eject 

\end{document}